\documentclass[conference]{IEEEtran}
\IEEEoverridecommandlockouts
\usepackage{cite}
\usepackage{url}
\usepackage{amsmath,amssymb,amsfonts}
\usepackage{algpseudocode}
\usepackage{graphicx}
\usepackage{textcomp}
\usepackage{xcolor}
\def\BibTeX{{\rm B\kern-.05em{\sc i\kern-.025em b}\kern-.08em
    T\kern-.1667em\lower.7ex\hbox{E}\kern-.125emX}}
\usepackage[utf8]{inputenc}
\usepackage{nomencl}
\makenomenclature

\begin{document}

\title{A Receding Horizon Reinforcement Learning Framework for Campus Chiller Energy Management - A case study from an Australian University\\
\thanks{* = Equal Contribution. Under review.}
}

\author{\IEEEauthorblockN{1\textsuperscript{st} Laura Musgrave *}
\IEEEauthorblockA{\textit{School of EECS} \\
\textit{The University of Queensland}\\
Brisbane, Australia \\
laura.musgrave@student.uq.edu.au}
\and
\IEEEauthorblockN{2\textsuperscript{nd} Arnab Bhattacharjee *}
\IEEEauthorblockA{\textit{School of EECS} \\
\textit{The University of Queensland}\\
Brisbane, Australia \\
a.bhattacharjee@uq.edu.au}
\and
\IEEEauthorblockN{3\textsuperscript{rd} Tapan Kumar Saha}
\IEEEauthorblockA{\textit{School of EECS} \\
\textit{The University of Queensland}\\
Brisbane, Australia \\
saha@eecs.uq.edu.au}
}

\maketitle

\begin{abstract}
This work presents a case study of optimal energy management of a large Heating Ventilation and Cooling (HVAC) system within a university campus in Australia using Reinforcement Learning (RL). The HVAC system supplies to nine university buildings with an annual average electricity consumption of $\sim2$ GWh. Updated chiller Coefficient of Performance (COP) curves are identified, and a predictive building cooling demand model is developed using historical data from the HVAC system. Based on these inputs, a Proximal Policy Optimization based RL model is trained to optimally schedule the chillers in a receding horizon control framework with a priority reward function for constraint satisfaction. Compared to the traditional way of controlling the HVAC system based on a reactive rule-based method, the proposed controller saves up to 28\% of the electricity consumed by simply controlling the mass flow rates of the chiller banks and with minimal constraint violations.  
\end{abstract}


\section{Introduction}

University campuses function as microcosms of urban energy systems, comprising diverse and interconnected energy infrastructures. A significant share of their total electricity consumption—often exceeding $40\%$—is attributed to Heating, Ventilation, and Air Conditioning (HVAC) systems. Within HVAC systems, central chiller plants dominate energy use, typically accounting for nearly $75\%$ of total HVAC electricity consumption. Optimizing their operation is therefore critical to achieving campus-wide energy efficiency targets and supporting broader carbon reduction goals.\\
A campus chiller plant usually consists of multiple chillers of varying rated capacities and part-load characteristics, along with variable-frequency drive (VFD) pumps that regulate the chilled-water flow rate. Such systems are intentionally over-provisioned to ensure reliability and cooling redundancy across several buildings. However, when operated using conventional rule-based or schedule-driven methods, they frequently encounter part-load inefficiencies, where chillers operate far from their optimal efficiency points. This can result in up to $10–20\%$ additional energy consumption compared to optimal operation, underscoring the need for intelligent control strategies capable of coordinating multi-chiller systems under dynamic cooling demands. Moreover, planning-based optimal control can also facilitate the operation of such manageable load sources for ancillary market participation, thereby enhancing grid stability and revenue generation\cite{10019581}. \\
Several control philosophies have been proposed for chiller plant optimization and can be broadly classified into three categories: rule-based, model-based, and data-driven approaches. Rule-based methods are the most prevalent in practice, relying on pre-defined load thresholds to sequentially switch chillers on or off in response to recent cooling demands \cite{JABARI2020106550}. While simple to implement, they fail to account for varying chiller efficiencies across part-load conditions, often resulting in sub-optimal operation. Model-based controllers such as Model Predictive Control (MPC) \cite{5530468,10867226} leverage physics-based or data-driven models of chiller dynamics and predictive load models to compute optimal control actions. Although MPC offers interpretability and explicit constraint handling, it is highly sensitive to model inaccuracies and typically requires convex relaxations to yield tractable optimization solutions \cite{6899700}.\\
The emergence of data-driven and learning-based control has enabled more adaptive HVAC management without explicit physical models. Deep reinforcement learning (DRL) and its multi-agent extensions have shown promise in learning optimal chiller control policies directly from operational data. For instance, \cite{HE2023107158} utilized an LSTM-based load predictor combined with a Deep Q-Learning (DQL) controller to regulate chiller supply temperature. Similarly, \cite{Manoharan2021} integrated meta-learning to improve sample efficiency and transferability across building environments, while \cite{luo2022controlling} demonstrated the potential of RL in commercial chiller operation through a combination of offline statistical analysis and online learning. Despite these advances, most existing studies treat all chillers as identical, neglecting heterogeneity in part-load efficiency characteristics that significantly influence overall plant performance and also struggle with physical constraint satisfaction. Furthermore, cooling demand forecasts in these works often rely solely on historical load data, omitting key exogenous factors such as ambient temperature and humidity, which limits generalization under varying weather conditions. As a result, reported energy savings typically plateau at around $10–12\%$, leaving considerable room for improvement.

This work proposes a deep reinforcement learning-based optimization framework for efficient operation of a campus chiller plant that supplies cooling to nine buildings in a large Australian university. The framework integrates predictive modelling and control to enhance both energy efficiency and adaptability. The key contributions of this study are summarized as follows:
\begin{enumerate}
    \item \textbf{Reinforcement-learning-based control}: A Proximal Policy Optimization (PPO) \cite{schulman2017proximal} agent is developed to regulate the chilled-water mass flow rates of a four-chiller plant. The agent operates in a receding-horizon manner, using 24-hour ahead predictions to plan control actions, of which only the first is executed at each step.
    \item \textbf{Reward function design with chiller heterogeneity}: A prioritized reward structure is formulated to promote hard constraint satisfaction and incorporate the distinct part-load efficiency profiles of individual chillers, guiding the RL agent toward energy-optimal operation.
    \item \textbf{Cooling demand forecasting}: A transformer-based sequence model, TimeXer, is employed to forecast building-level cooling demand using both historical demand and weather variables, enhancing predictive accuracy under dynamic conditions.
    \item \textbf{Performance benchmarking}: The proposed receding-horizon PPO agent is benchmarked against a reactive load-based baseline controller and a standard single-stage RL controller. Experimental results demonstrate up to $28\%$ power savings compared to the rule-based baseline and $8\%$ savings relative to the non-receding RL configuration.
\end{enumerate}

By integrating domain-informed reward design with data-driven forecasting and receding-horizon learning, this work demonstrates a scalable approach to energy-efficient chiller plant control that bridges the gap between traditional model-based optimization and modern reinforcement learning.

\section{Problem Formulation}
\begin{figure}[htbp]
\centering
\includegraphics[width=0.95\columnwidth]{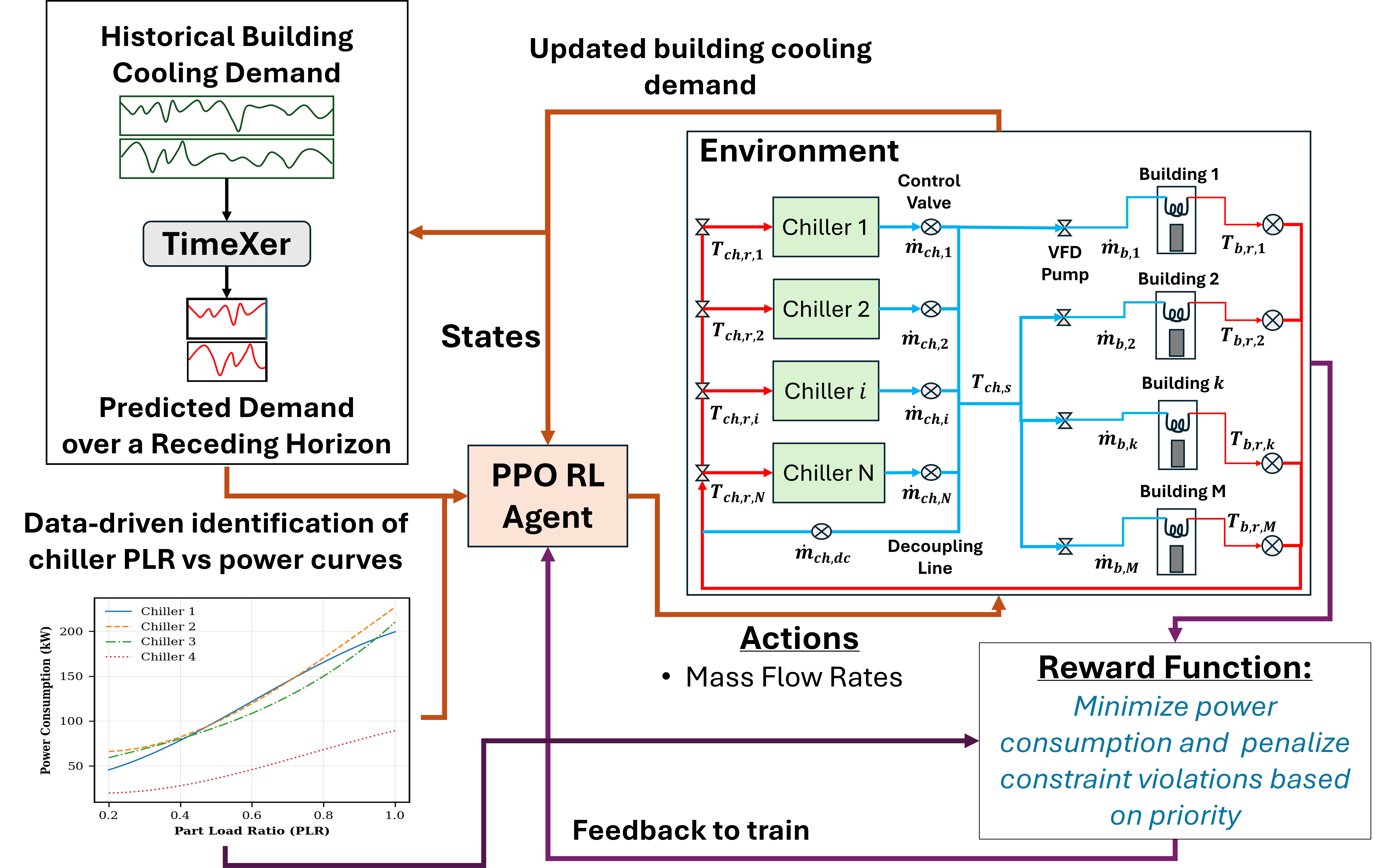 }
\caption{A pictorial representation of a campus chiller network supplying to multiple buildings.}
\label{fig:RL_Framework}
\end{figure}

A receding horizon controller for predictive chiller load management is designed using a PPO-based RL agent. At every control instant, a building cooling load prediction algorithm generates possible building load trajectories over a 24 hour horizon which is used by the RL agent to schedule the chillers optimally by solving a constrained power minimization problem over a rolling horizon. To specify the differences in the part load efficiencies of the chillers, part-load ratio (PLR)-vs-Power consumption curves are pre-determined from historical data of the chillers and are explicitly incorporated in the reward function used to train the agent. An overview of the proposed RL framework is given in Fig.\ref{fig:RL_Framework} with the chiller network depicted as the environment. Specifically, the RL agent controls the mass flow rates of the chillers due to the following reasons: 
\begin{enumerate}
    \item Mass flow rates provide a wider range of control compared to supply temperatures and are controllable in the considered system through variable frequency motor systems.
    \item The existing rule-based control strategy of the chillers fixes the supply temperature at $\sim 6^o C$ during working hours and $ \sim 11^o C$ during non-working hours while varying the mass flow rates of the 'ON' chillers reactively based on load demand.
\end{enumerate}
However, the authors acknowledge that this strategy may not be applicable in constant-flow chiller systems and will continue exploration of selective control of supply temperature and mass flow rates as part of future research. 

\subsection{Assumptions}
In line with existing literature\cite{5530468,10867226, HE2023107158, Manoharan2021}, the following assumptions were made when formulating the optimization problem:
\begin{itemize}
\item The reticulation pipework is lossless (i.e. the supply water temperature from the chillers is the same as the supply water temperature to the buildings).
\item The total mass flow rate from the chillers is equal to the total mass flow rate from the buildings.
\item The supply water temperature for the chillers is fixed at $6^o$C when the they are in operation.
\item The return water temperature is equal for all chillers.
\end{itemize}

\subsection{Objective Function}
Consider a chiller bank with $n_c$ chillers of varying cooling capacities $\dot{Q}_{r,i}$. The intended goal is to set optimal mass flow rates for the chillers such that the power consumption of the chiller bank is minimized over a $N$ step prediction horizon. This can be expressed mathematically by,

\begin{equation}
\min \sum_t^{t+N \cdot \Delta{t}} \left(\sum_{i=1}^{n_c} P_{c,i}\right)
\end{equation}

A receding horizon approach is used, in which the optimization problem is solved after every $\Delta t$ in the control horizon subject to the constraints outlined in the following sections.

\subsection{Main Constraints}\label{section: constraints}

\subsubsection{Power Consumption Equation}
$P_{c,i}$ is the power consumed by chiller $i$ over {[$t, t+\Delta t$]}, and can be represented by the equation,

\begin{equation}\label{eqn: power curve}
P_{c,i} = \alpha_i + \beta_i \cdot PLR_i + \gamma_i \cdot PLR_i^2 + \psi_i \cdot PLR_i^3
\end{equation}

where $\alpha_i, \beta_i, \gamma_i, \psi_i$ are constant coefficients (see section \ref{section: power consumption model} for details), and $PLR_i$ is the chiller Part Load Ratio.

\subsubsection{PLR Equation} 
The PLR is the ratio of a chiller's cooling output $\dot{Q}_{c,i}$ to it's rated cooling capacity $\dot{Q}_{r,i}$, and is given by,

\begin{equation}
PLR_i = \frac{\dot{Q}_{c,i}}{\dot{Q}_{r,i}}
\end{equation}

\subsubsection{Chiller Cooling Equation}
The amount of cooling $\dot{Q}_{c,i}$ provided by chiller $i$ is given by,

\begin{equation}\label{Q_dot}
\dot{Q}_{c,i} = \delta_i \dot{m}_{c,i}c_{w}(T_{cr}-T_{cs})
\end{equation}

where $\dot{m}_{c,i}$ is the chiller mass flow rate, $c_w$ is the specific heat capacity of water, $T_{cr}$ and $T_{cs}$ represent the return and supply water temperature respectively, and $\delta_i$ is a binary decision variable,

\begin{equation}\label{eqn:chiller set}
\delta_i = \left\{
    \begin{aligned}
        & \text{1, if chiller $i$ is $ON$} \\
        & \text{0, if chiller $i$ is $OFF$}
    \end{aligned}
    \right.
\end{equation} 

When a chiller is $ON$, it supplies chilled water at varying mass flow rates and a fixed supply temperature of approximately $6^o$C. When $OFF$, it is considered shut down and supplies no chilled water. Thus, for a given chilled water return temperature $T_{cr}$, control of the cooling output (and by extension, the power consumption) of the chiller $i$ can be achieved by varying it's mass flow rate $\dot{m}_{c,i}$.

\subsubsection{Chilled Water Return Temperature}
The temperature of the chilled water returning to the chiller bank $T_{cr}$ varies according to the system dynamics (i.e. the heat retention characteristics of the building and the total thermal load). Given the assumption that the pipes are lossless, the return water temperature for the chillers can be expressed as,

\begin{equation}\label{eqn:t_return}
T_{cr} = \frac{\dot{Q}_b}{\left(\sum_{i=1}^{n_c}\delta_i \dot{m}_{c,i}\right)c_w} + T_{cs}
\end{equation}

\subsubsection{Energy Balance Equation}
The system energy balance is given by,
\begin{equation}\label{eqn: energy balance}
\sum_{i=1}^{n_c} \dot{Q}_{c,i} = \dot{Q}_{b}
\end{equation}

\subsubsection{PLR Operational Range}
When chiller $i$ is $ON$, it's PLR is subject to the constraint,
\begin{equation}\label{eqn: min PLR}
PLR_{min} \leq PLR_i \leq PLR_{max}
\end{equation}

\subsubsection{Return Temperature Operational Range}
When any of the chillers are $ON$, the return temperature of the chiller bank is subject to the constraint,
\begin{equation}\label{eqn: Tr constraint}
T_{{cr}_{min}} \leq T_{cr} \leq T_{{cr}_{max}}
\end{equation}

\subsubsection{Minimum Operational Mass Flow Rate}
When chiller $i$ is $ON$, it's mass flow rate is subject to the constraint,
\begin{equation}\label{eqn: mfr contraint}
\dot{m}_{{c,i}_{min}} \leq \dot{m}_{c,i}
\end{equation}

\subsubsection{Maximum Coefficient of Performance}
The Coefficient of Performance (COP) of chiller $i$ is given by $COP_i = \frac{\dot{Q}_{c,i}}{P_{c,i}}$. The highest COP for chiller $i$ is given by,
\begin{equation}\label{eqn: cop constraint}
COP_i \leq COP_{i_{max}}
\end{equation}

\section{Optimal Control Strategy}
\subsection{Reinforcement Learning for Optimal Chiller Scheduling}
 A Proximal Policy Optimization based RL controller is proposed to learn a scheduling policy that selects the mass flow rates and, by extension, the $ON/OFF$ status of the chillers.

\subsubsection{State-space} The state vector $s_t$ for the agent is given by,
\begin{equation}
s_t = \{\hat{\dot{Q}}_b, \dot{Q}_{c,1...n_c}, PLR_{c,1...n_c}, COP_{1...n_c}, P_{c,1...n_c}, T_{cr}\}_t
\end{equation}
The building cooling load is given by predicted building cooling load, $\hat{\dot{Q}}_b$ in the state vector.

\subsubsection{Action-space} The action space is given by,
\begin{equation}
a_t = \{m_{c,1...n_c}\}_t
\end{equation}

\subsubsection{Reward}
The reward function was developed with the aim of promoting constraint satisfaction while achieving the goal of minimizing power consumption. A sequential structure was used, implicitly assigning a priority to the constraints as the agent would receive a large negative reward until each constraint was met. Formally, suppose each constraint $C_i \forall i \in |\mathcal{C}|$ where $\mathcal{C}$ is the total constraint set, corresponds to a reward component $r_i$, then the reward function at step $t$ is modelled as:
\begin{equation}
    r_t = \sum_{c=1}^{|\mathcal{C}|} \lambda_c r_{c,t} 
\end{equation}
where $\sum_{c=1}^{|\mathcal{C}|} |\lambda_c| = max(\{|\lambda_c|\}_{c=1}^{|\mathcal{C}|})$ implies that any one of the reward components is active at time $t$.  A pre-determined ordering strategy is used based on expert knowledge to select which constraints violations are more serious than others. Suppose, the priority is given by $f_P: \mathcal{C} \to \mathbb{N}$ such that $n =  f_P(C_k)$ implies that the constraint $C_k$ is the $n^{th}$ constraint to be satisfied in order, then the following condition determines the value of the reward function at any time $t$ : $|\lambda_k| > 0$ iff $C_k$ is violated and $\forall i \in \mathcal{C}$ such that $f_P(C_i) < f_P(C_k)$, $C_i$ is not violated. In this ordering, the reward component that corresponds to the power consumption is the least prioritized, ensuring that the agent focuses on meeting the hard constraints prior to minimizing power consumption. The hard constraints considered in this work are the energy balance equation (Eq. \ref{eqn: energy balance}) and satisfying the return temperature bounds (Eq. \ref{eqn: Tr constraint}) as they were found to be the hardest to satisfy during experiments. Only once the agent's action meets all these  hard constraints can a positive reward be obtained. It should be noted that negative behaviours from the agent are penalized more heavily than positive behaviours are rewarded. The remaining constraints are considered soft constraints and included in the base reward with varying weights. Auxiliary soft constraints are included in the base reward to minimize frequent chiller switching and promoting sparsity in the number of operational chillers.

\subsection{Building Cooling Load Forecasting}
The receding horizon approach taken by the RL agent relies on the accurate forecasting of the building cooling load over the prediction horizon $N$. The TimeXer~\cite{wang2024timexer} module was selected for this task due to its ability to capture intra-endogenous temporal dependencies as well as exogenous-to-endogenous correlations in its forecasting approach . 

The exogenous variables used in the implementation of the forecaster include average wind speed and direction, average ambient temperature, relative humidity, and average solar irradiance amount. The trained forecaster produces multivariate predictions of length 48 (i.e. 24 hours) using a sequence length of 336 (i.e., 7 days).

\section{Experiment Setup}

\subsection{Overall System Description}
The chilled water refrigeration plant considered in this study is comprised of four water cooled chillers with three of them rated at 1700 kWr and one at 710 kWr, four cooling towers, chilled water and condenser water pumps, and associated pipework, valves and accessories. The upper limits of the mass flow rates for Chillers 1 to 3 and Chiller 4 are 50.8 L/s and 21.2 L/s respectively. This chilled water plant serves up to nine campus buildings at once via pipework which reticulates chilled water throughout the buildings to air handling units (AHUs), fan coils, and under-ceiling fan coils. The lower and upper limits for different parameters are as follows: $PLR_{min} = 0.2$, $PLR_{max} = 10$, $T_{cr,min} = 6.56$, $T_{cr,max} = 14$, $\dot{m}_{c,min} = 14.5$ and $COP_{max} = 10$ for Chillers 1 to 3 and $COP_{max} = 5.814$ for Chiller 4. 

The RL agent utilizes the OpenAI SpinningUp PPO algorithm~\cite{spinningup2018}. Training took approximately 3 hours using the Goolge Colab T4 GPU Runtime (maximum GPU RAM of 15.0 GB)



\subsection{Chiller Power Consumption Model}\label{section: power consumption model}

Because the reward function incorporates chiller power consumption, a method of predicting this value for a given chiller cooling output was required. The degree three polynomial in Eq. \ref{eqn: power curve} was used in the proposed algorithm, where the values for coefficients $\alpha_i$, $\beta_i$, $\gamma_i$ and $\psi_i$ were found by performing a polynomial regression on historical data for each chiller. The results of the regression are summarised in Table \ref{tab: power curve coeffs} below.

\begin{table}[htbp]
\caption{Coefficients for the Chiller Power Curves}
\begin{center}
\begin{tabular}{|c|c|c|c|c|c|}
\hline
\textbf{Chiller No.} & {$\boldsymbol{\alpha}$}& {$\boldsymbol{\beta}$}& {$\boldsymbol{\gamma}$}&
{$\boldsymbol{\Phi}$} & $\boldsymbol{R^2}$ \\
\hline
1&33.3469&-7.3826&384.8817&-211.3766&0.9332\\
\hline
2&78.6233&-147.0632&454.0039&-158.7085&0.8699\\
\hline
3&39.5435&103.8645&-53.5269&120.3910&0.9120\\
\hline
4&27.0384&-115.1061&261.4245&-84.0566& - \\
\hline

\end{tabular}
\label{tab: power curve coeffs}
\end{center}
\end{table} 

The absence of an $R^2$ value for Chiller 4 is because there was not enough historical data to produce a statistically significant trend. As such, it's curve was derived from data in the mechanical specification.

\section{Results}

\subsection{Building Load Forecast}
\begin{figure}[htbp]
\centering
\includegraphics[width=\columnwidth]{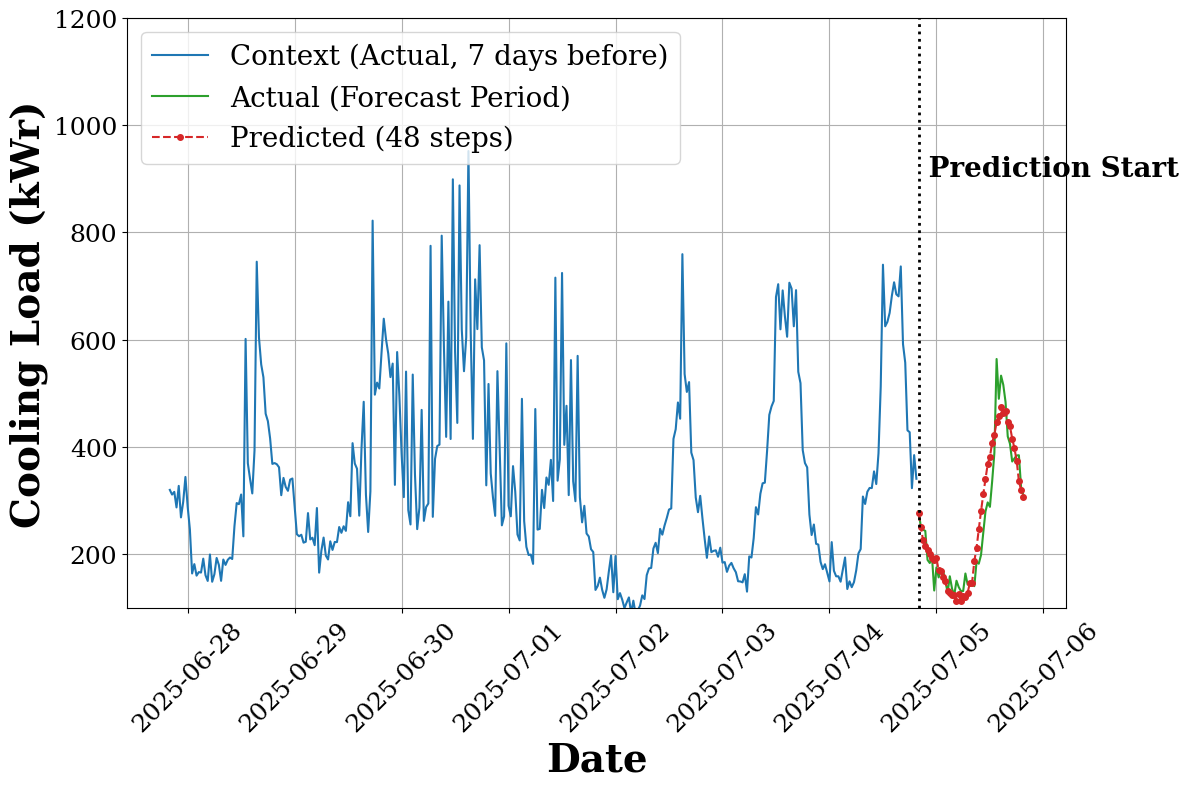}
\caption{Building load forecast with TimeXer}
\label{fig:building_load_forecast}
\end{figure}
Fig. \ref{fig:building_load_forecast} shows the building forecast results using the TimeXer module for a specific testing instant. The normalized mean absolute error obtained for the testing period is $0.235$, which results in quite accurate forecasts. 
\subsection{Benchmarking Results}
\begin{figure}[htbp]
\centering
\includegraphics[width=0.6\columnwidth]{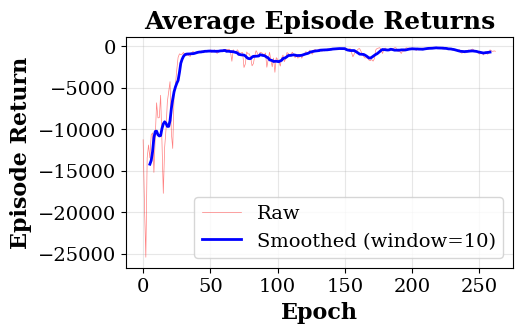}
\caption{Training results for the RL agent.}
\label{fig:constraints_returns}
\end{figure}

The training time return curve for the RL agent is shown in Fig. \ref{fig:constraints_returns}. As training progresses, the return values become non-zero, implying satisfaction of the hard constraints. 
The power savings from using the proposed RL is compared against two baselines - the real-life rule-based controller and a pure RL controller that considers just the most recent power consumption in the state vector and does a one-step control instead of a receding horizon control. All three methods are evaluated over a two month evaluation period and the results are shown in Table \ref{tab:power_savings}. The proposed RL-based receding horizon controller results in a total energy consumption of $66.49~MWh$, compared to $ 92.35~MWh$ using the rule-based baseline and $74.35~MWh$ using the pure RL controller, resulting in a $28\%$ savings in power consumption from the rule-based baseline applied in practice. These savings were achieved through two primary means: improving the accuracy of matching cooling demand and supply, and improving the COP of operating chillers. 

\begin{table}[htbp]
\centering
\caption{Energy Savings comparison across rule based and RL strategies. RH = Receding Horizon}
\label{tab:power_savings}
\resizebox{\columnwidth}{!}{%
\begin{tabular}{|l|c|c|c|}
\hline
\multicolumn{1}{|c|}{\textbf{Control Method}} & \textbf{Energy Consumed (MWh)} & \textbf{Energy Saved (MWh)} & \textbf{\% Energy Saved} \\ \hline
 
\textbf{Rule-Based (in practice)}                   & 92.35                          & 0                           & 0                        \\ \hline
 
\textbf{Pure RL (no RH)}              & 74.15                          & 18.2                        & 19.7                     \\ \hline
 
\textbf{Proposed RL}                  & \textbf{66.49}                 & \textbf{25.86}              & \textbf{28}              \\ \hline
\end{tabular}
}
\end{table}


\subsection{Constraint Satisfaction}
\begin{figure}[htbp]
\centering
\includegraphics[width=\columnwidth]{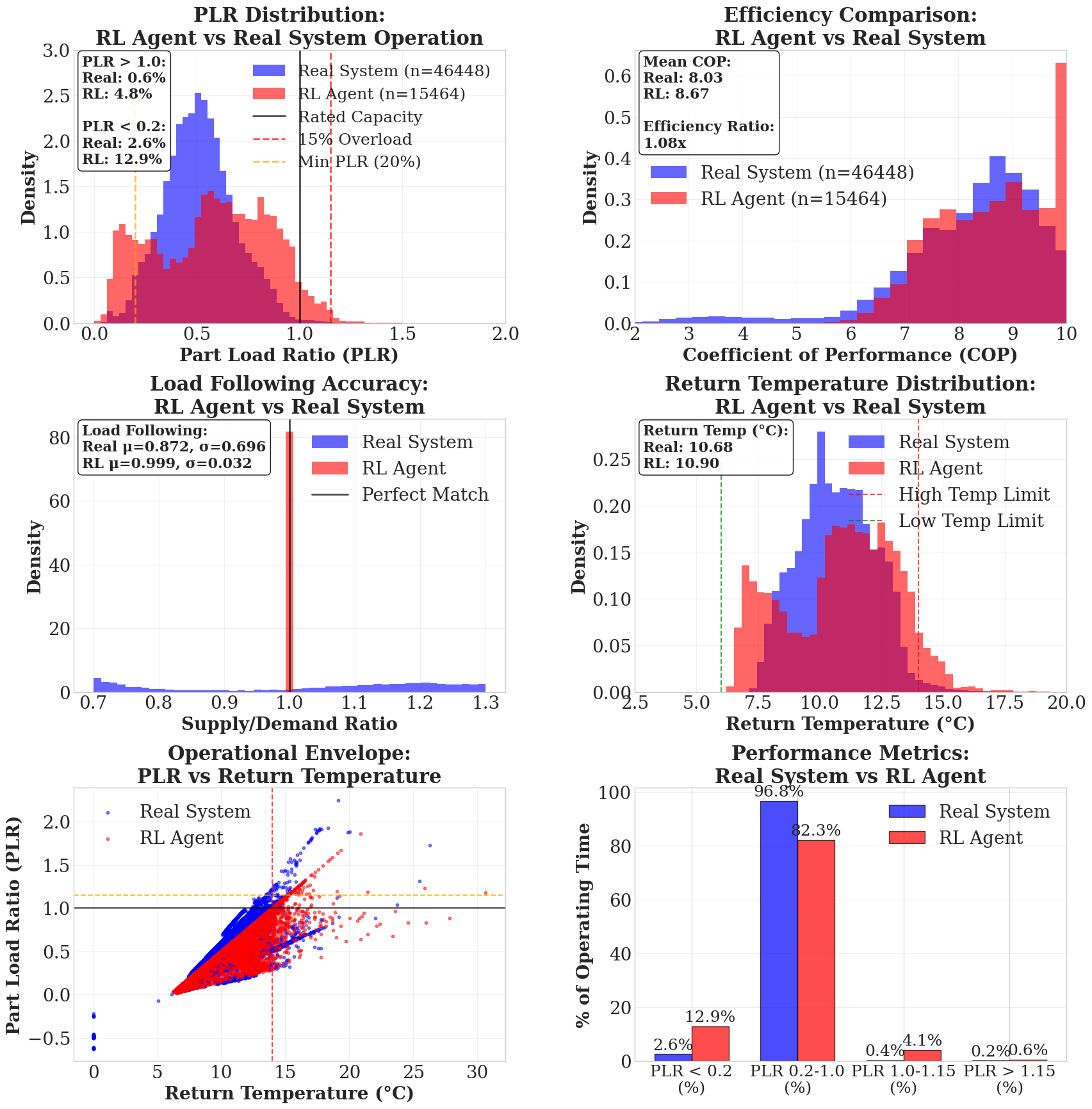}
\caption{Summary of results from the controller evaluation.}
\label{fig:results_summary}
\end{figure}

Fig. \ref{fig:results_summary} compares constraint satisfaction performance between a baseline rule-based controller (blue) and the proposed reinforcement learning (RL) agent (red) for chiller bank operation. The RL agent achieves higher mean Part Load Ratio (0.83 vs. 0.81) and Coefficient of Performance (8.81 vs 8.03), indicating improved efficiency. Temperature distributions are comparable, with both models maintaining thermal constraints. Load-following accuracy is significantly better for the RL agent than the baseline. The RL agent apparently violates the lower limit of the PLR on more instances than the baseline, but these happen only for short periods of time when the chilling load requirement is significantly low $(< 300 kWr)$. This limitation will however be addressed in subsequent research. Overall these results demonstrate the proposed RL model’s superior constraint adherence and operational efficiency across multiple metrics while being able to significantly reduce energy consumption.


\section{Conclusion}
This paper proposes a reinforcement learning framework for optimal chiller scheduling in campus HVAC systems, combining transformer-based cooling demand forecasting with a receding horizon PPO agent. The controller achieves up to $28\%$ electricity savings while minimizing constraint violations. However, assumptions like lossless pipework and manual constraint prioritization may limit scalability. The forecaster also lacks real-time retraining. Future work will address these limitations by modeling pipe losses, automating reward shaping, enabling online learning, and extending the framework for demand response and ancillary service participation using HVAC systems.

\bibliographystyle{unsrt}
\bibliography{Bibliography.bib}

\end{document}